\documentclass[twocolumn,floats,amsmath,amssymb,preprintnumbers,superscriptaddress,floatfix,prl]{revtex4-1}

\usepackage{graphicx}
\usepackage{bm}
\usepackage{color}
\usepackage{hyperref}
\usepackage{ulem}
\usepackage{amsmath}
\usepackage{amssymb}
\usepackage{mathtools}
\usepackage{mathrsfs}

 \usepackage{here}
 \usepackage[dvipsnames]{xcolor}

\newcommand{\kB}{k_\mathrm{B}}

\newcommand{\kT}{k_\text{B} T}

\def\Eq{Eq.}

\def\Figure{Figure}

\def\Fig{Fig.}

\extrafloats{100}

\begin{document}

\title{Tuning the permeability of dense membranes by shaping nanoscale potentials}

\author{Won Kyu Kim}
\email{wonkyu.kim@helmholtz-berlin.de}
\affiliation{Research Group for Simulations of Energy Materials, Helmholtz-Zentrum Berlin, D-14109 Berlin, Germany}

\author{Matej Kandu\v{c}}
\affiliation{Research Group for Simulations of Energy Materials, Helmholtz-Zentrum Berlin, D-14109 Berlin, Germany}
\affiliation{Jo\v{z}ef Stefan Institute, SI-1001 Ljubljana, Slovenia}

\author{Rafael Roa}
\affiliation{F\'isica Aplicada I, Facultad de Ciencias, Universidad de M\'alaga, 29071 M\'alaga, Spain}

\author{Joachim Dzubiella}
\email{joachim.dzubiella@physik.uni-freiburg.de}
\affiliation{Research Group for Simulations of Energy Materials, Helmholtz-Zentrum Berlin, D-14109 Berlin, Germany}
\affiliation{Applied Theoretical Physics-Computational Physics, Physikalisches Institut, Albert-Ludwigs-Universit\"at Freiburg, D-79104 Freiburg, Germany}

\date{\today}

\begin{abstract}

The permeability is one of the most fundamental transport properties in soft matter physics, material engineering, and nanofluidics. 
Here we report by means of Langevin simulations of ideal penetrants in a nanoscale membrane made of a fixed lattice of attractive interaction sites, how the permeability can be massively tuned, even minimized or maximized,  by tailoring the potential energy landscape for the diffusing penetrants, depending on the membrane attraction, topology, and density.  
Supported by limiting scaling theories we demonstrate that the observed non-monotonic behavior and the occurrence of extreme values of the permeability is far from trivial and triggered by a strong anti-correlation and substantial (orders of magnitude) cancellation between penetrant partitioning and diffusivity, especially within dense and highly attractive membranes. 
\end{abstract}


\maketitle

Permeability defines the ability of penetrating molecules (e.g., gas, ligands, reactants, etc.) to collectively permeate and flow through a given  medium under the action of an external field or chemical gradient.  It is thus without doubt one of the most fundamental transport descriptors employed in the physical sciences and material engineering.  In the standard `solution--diffusion' picture for dense membranes it is commonly defined on the linear response level  by~\cite{yasuda1969permeability3,robeson, Baker, gehrke, chauhan,Thomas2001transport,Ulbricht2006, Baker2014,park2017} 
\begin{equation}\label{eq:P}
\mathcal{P} = \mathcal{K} D_{\rm in},
\end{equation}
where $\mathcal{K} = c_{\rm in}/c_{\rm 0}$ is the equilibrium partitioning defined as the ratio of number densities of the penetrants inside and outside the medium, and $D_\text{in}$ is the diffusion coefficient of those inside. The optimization of permeability, especially for being highly selective among different penetrants, has been a grand challenge in material design over the last decades~\cite{prevost:2007, chauhan, park2017}. Prominent applications revolve around gas separation or recovery~\cite{robeson, chauhan, MOF, lyd, obliger, park2017}, desalination and nanofiltration (`molecular sieving') ~\cite{shannon,geise2011,tansel}, medical treatments by dialysis or selective drug transport~\cite{peppas1999, stamatialis}, or hydrogel-based soft sensors or nanoreactors~\cite{palasis1992permeability, stuartNatMat2010, Lu2011, Rafa2017}. The membrane materials range from solid nanoporous carbon or silica to metal organic frameworks to soft polymer matrices, for all of which the topology and chemistry can be well controlled and fine-tuned nowadays. 

\begin{figure}[b]
\centering
\includegraphics[width = 0.35\textwidth]{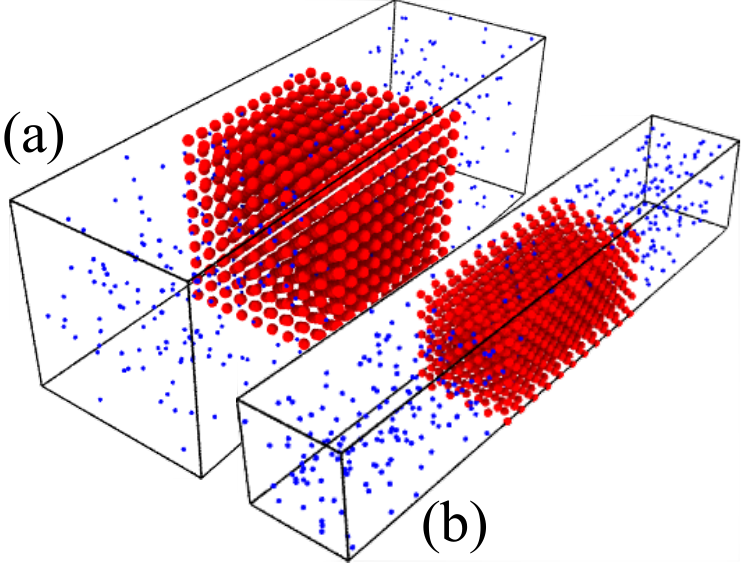}
\caption{Snapshots of the two simulated membrane--penetrant systems. 
The interaction sites in the membrane (red) are fixed on a lattice, and the penetrants (blue) are diffusing and interacting with the membrane sites via the LJ potential. Two different lattices are considered:~(a)~Simple-cubic (SC) and (b)~Face-centered-cubic (FCC) lattices.
}\label{fig:fig1}
\end{figure}

Theoretical attempts to model permeability have started mostly with simple `free volume' or `obstruction' theories for both partitioning and diffusion~\cite{Yasuda1968,yasuda1969permeability2,yasuda1969permeability3,robeson, Baker, gehrke, masaro1999physical, Amsden1998}. It has turned out, particularly with the help of computer simulations, that the details are much more complex due to the various specific molecular interactions and topologies inside the membranes: on one hand, the partitioning, that is, `solvation' of molecular penetrants in the dense media in general results from a competition between various, e.g., steric, solvophobic, dispersion, and electrostatic potentials~\cite{Ben, MOF, moncho2014ion, lyd, obliger,  adroher2015role,Erbas2016,Ben2,kim2017cosolute}. It was shown recently that this competition can lead to a maximization of partitioning of penetrants in polymer membranes tuned by volume fraction~\cite{monchoPCCP2018}. On the other hand,  diffusion in dense membranes is highly non-viscous and qualitatively length-scale and potential dependent~\cite{lyd,obliger, cai2015hopping, Ben2, zhang2017molecular,  zhang2017correlated,Huskey2016, Matej2018macro}. In particular, increasing attraction of the penetrants was shown to lead to strikingly non-monotonic diffusion, featuring massive slow-downs in dense membranes due to trapping~\cite{hansing2016nanoparticle, Putzel2014a,Ghosh2014a}. While partitioning and diffusion have thus received much attention individually, no systematic study exists on their combined impact on the product $\mathcal{P}$.  

\begin{figure*}[t]
\centering
\includegraphics[width = 0.8\textwidth]{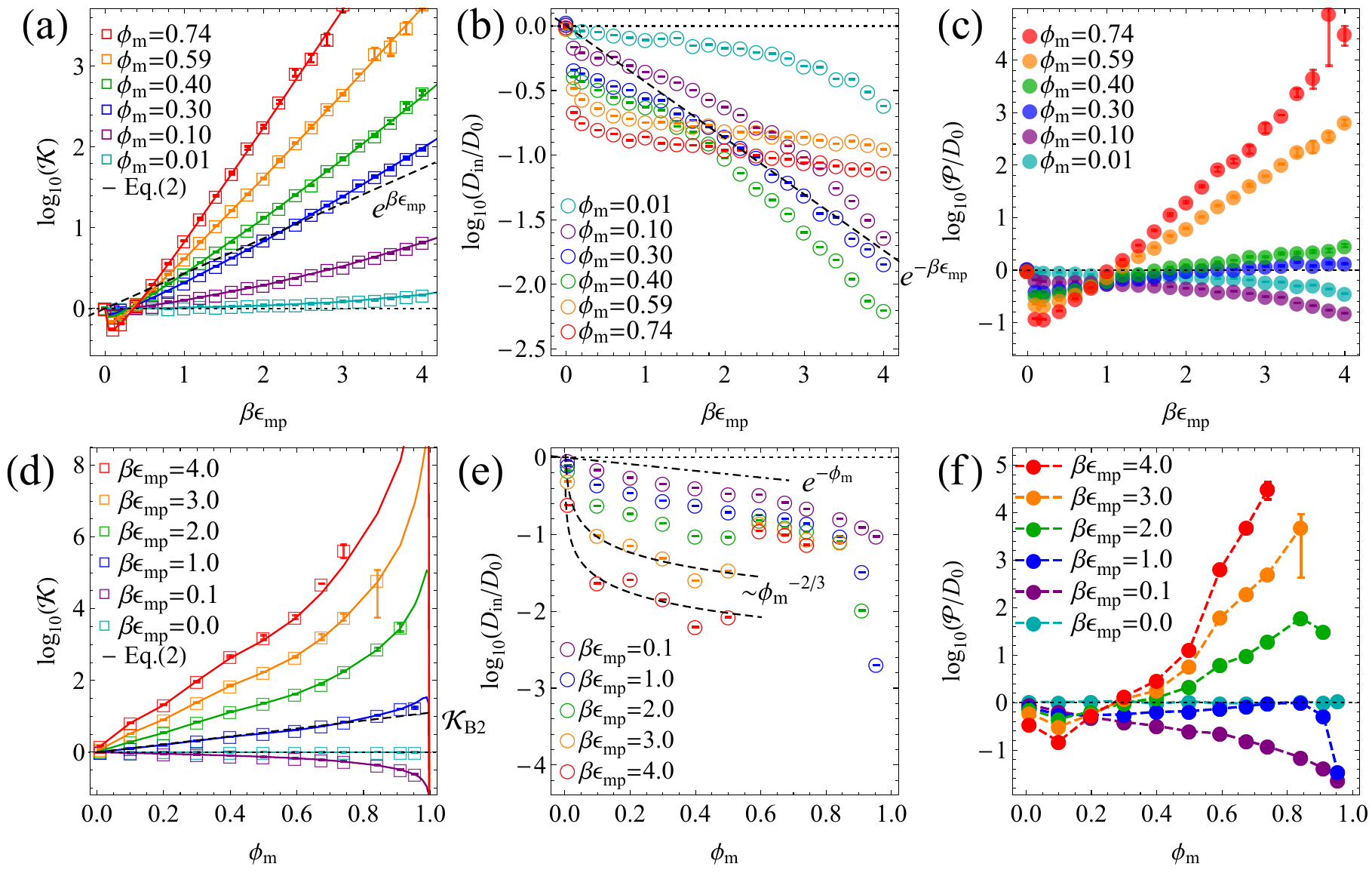}
\caption{
Simulation results (symbols) for the simple-cubic (SC)  membrane--penetrant systems, depending on the membrane--penetrant LJ interactions $\epsilon_\text{mp}$ and the membrane volume fractions $\phi_\text{m}$.
(a)~Penetrant partitioning $\mathcal{K}(\epsilon_\text{mp})$ at different $\phi_\text{m}$. The solid lines depict the exact relation given by \Eq~\eqref{eq:K}. 
The dashed line shows the scaling $\mathcal{K} = e^{\beta \epsilon_\text{mp}}$.
(b)~Penetrant diffusivity $D_\text{in}(\epsilon_\text{mp})/D_0$ at different $\phi_\text{m}$. The dashed line depicts the scaling $D_\text{in} / D_0 = e^{-\beta \epsilon_\text{mp}}$.
(c)~Permeability $\mathcal{P}(\epsilon_\text{mp})/D_{0}=\mathcal{K} D_\text{in}/D_{0}$ at different $\phi_\text{m}$.
(d)~$\mathcal{K}(\phi_\text{m})$ at different $\epsilon_\text{mp}$. The solid lines depict the exact relation in \Eq~\eqref{eq:K}. The dashed line depicts the approximation $\mathcal{K}_\text{B2}(\phi_\text{m})$ at $\beta \epsilon_\text{mp} = 1$ (see text for details).
(e)~$D_\text{in}(\phi_\text{m}) / D_0$ at different $\epsilon_\text{mp}$. The dashed lines depict the approximation $D_\text{in} / D_0 = \exp (- \phi_\text{m})$ valid for low $\epsilon_\text{mp}$ and $\phi_\text{m}$, and the Kramers scaling $D_\text{in} / D_0 \sim \phi_\text{m}^{-2/3}$ with two different prefactors, valid for high $\epsilon_\text{mp}$ and low $\phi_\text{m}$.
(f)~$\mathcal{P}(\phi_\text{m})/D_{0}$ at different $\epsilon_\text{mp}$.
}\label{fig:KDP_SC}
\end{figure*}

In this letter we demonstrate using Langevin dynamics simulations of a minimalistic model system of ideal penetrants in a dense lattice membrane how the permeability can be tuned massively, even maximized or minimized, by systematically varying the attraction and volume fraction as well as the topology of the membrane. This nontrivial non-monotonicity results from a strong anti-correlation between penetrant partitioning and diffusivity, especially in attractive membranes.
Our findings thus provide design rules for synthetic membranes to optimize selectivity and performance of functional and nanofluidic transport devices.

\textit{Methods.}---
We perform Langevin dynamics simulations~\cite{LAMMPS} of membrane--penetrant systems as shown in \Fig~\ref{fig:fig1}: 
A long periodic simulation box is considered, where in the central slab of volume $V_\text{m}$ (the membrane) there is a simple-cubic (SC) or face-centered-cubic (FCC) lattice of the membrane-constituting molecules, i.e., spherical sites (red) whose positions are fixed with the lattice constant $l$. 
The penetrants (blue) are diffusive throughout the whole simulation box and interact with the sites via the Lennard-Jones (LJ) potential $U_\text{mp} (r) = 4 \epsilon_\text{mp} [(\sigma_\text{mp} / r)^{12} - (\sigma_\text{mp} / r)^{6}]$, where for the ideal point-like penetrants $\sigma_\text{mp} = r_\text{s}$, the radius of the site.  By varying $r_\text{s}$, we control the membrane volume fraction $\phi_\text{m} = v/ V_\text{m}$, where $v$ is the volume occupied by the sites. The radius $r_\text{s}$ can be larger than $l$, so we also allow overlapping between the sites. Details of the methods can be found in the Supplemental Material~\cite{SI}. 

We compute the permeability for various membrane volume fractions $\phi_\text{m}$ and membrane--penetrant interaction energies $\epsilon_\text{mp}$. 
The partitioning is obtained in equilibrium using $\mathcal{K} = c_{\text{in}}/c_{0}$, i.e., averaging the density of penetrants inside the slab.  
Generally, $\mathcal{K}$ for the ideal penetrants is defined via the excess chemical potential $\Delta \mu = -\kT \ln \overline{ e^{-\beta H_\text{mp}} }$ through $\mathcal{K} = e^{-\beta \Delta \mu}$~\cite{Leo1971}, where $\kT =1/\beta$ denotes the thermal energy, $H_\text{mp}(\mathbf r)=\sum_i U_{\rm mp}(|\mathbf r - \mathbf r_i|)$ is the membrane--penetrant interaction Hamiltonian (summing over all sites $i$), and $\overline{x} \equiv \int \text{d}V x / V_c$ denotes the slab volume average, yielding exactly
\begin{equation}
\label{eq:K} \mathcal{K}  =  \overline{ e^{-\beta H_\text{mp}} },
\end{equation}
enabling a direct comparison and verification of the simulation results.
\textcolor{black}{To compute the penetrant long-time self-diffusivity $D_\text{in}$, we perform additional simulations of penetrants in a periodic box of the lattices, and evaluate the mean-squared-displacement $\text{MSD} = 6 D_\text{in} t$ in the long time limit, ensuring normal diffusion in the overdamped regime~\cite{SI}.}

\begin{figure*}[t]
\centering
\includegraphics[width = 0.8\textwidth]{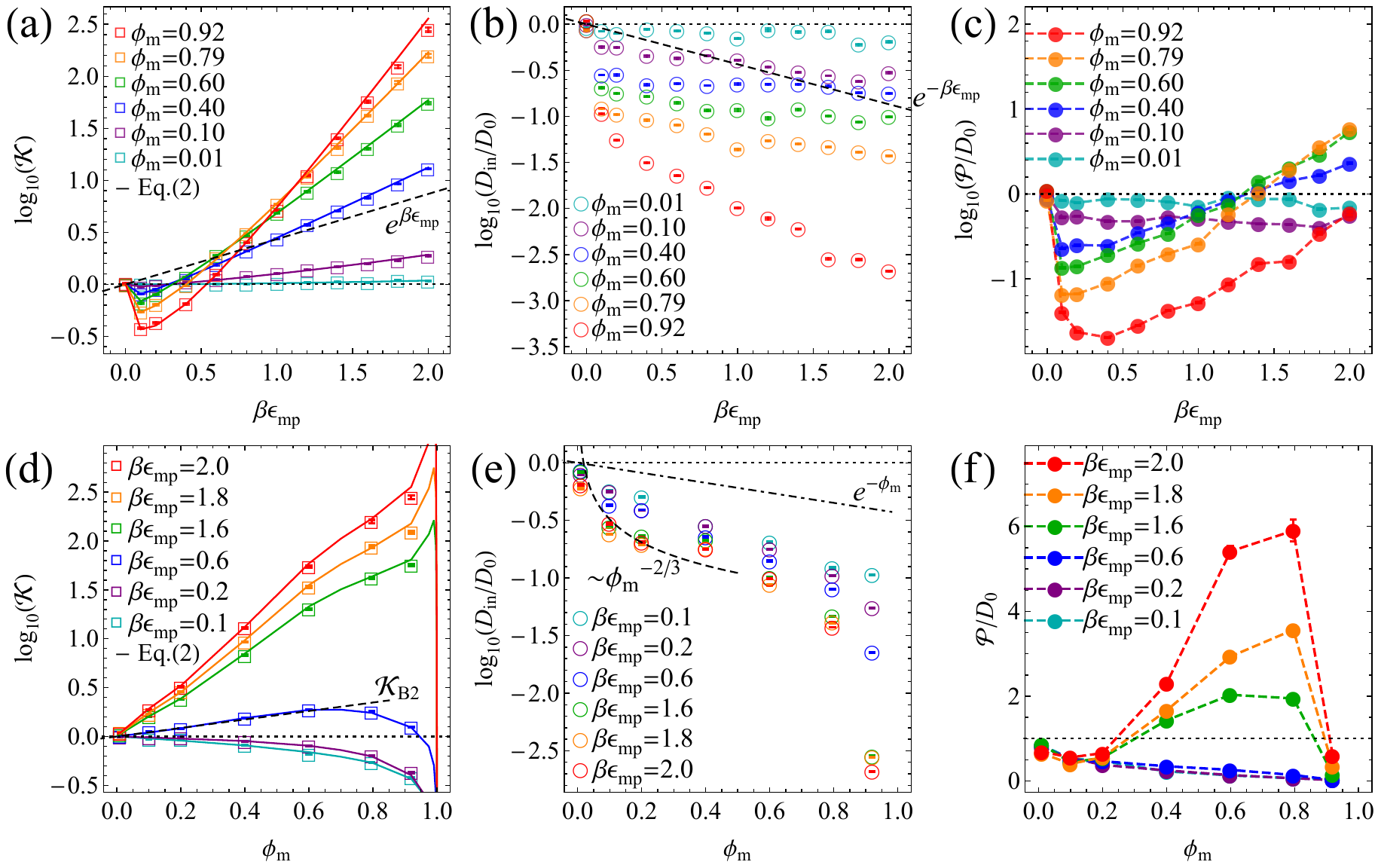}
\caption{
Simulation results (symbols) for the FCC membrane--penetrant systems, depending on the membrane--penetrant LJ energy $\epsilon_\text{mp}$ and the membrane volume fraction $\phi_\text{m}$. Note that we plot the decadic log of the observables. 
(a)~Penetrant partitioning $\mathcal{K}(\epsilon_\text{mp})$ at different $\phi_\text{m}$. The solid lines depict the exact relation in \Eq~\eqref{eq:K}. 
The dashed line shows the scaling $\mathcal{K} = e^{\beta \epsilon_\text{mp}}$.
(b)~Penetrant diffusivity $D_\text{in}(\epsilon_\text{mp})/D_0$ at different $\phi_\text{m}$.The dashed line depicts the scaling $D_\text{in} / D_0 = e^{-\beta \epsilon_\text{mp}}$.
(c)~Permeability $\mathcal{P}(\epsilon_\text{mp})/D_{0}=\mathcal{K} D_\text{in}/D_{0}$ at different $\phi_\text{m}$.
(d)~$\mathcal{K}(\phi_\text{m})$ at different $\epsilon_\text{mp}$. The solid lines depict the exact relation in \Eq~\eqref{eq:K}, and the dashed line depicts the approximated partitioning $\mathcal{K}_\text{B2}$ at $\beta \epsilon_\text{mp} = 0.6$ (see text for details).
(e)~$D_\text{in}(\phi_\text{m}) / D_0$ at different $\epsilon_\text{mp}$. The dashed lines depict the approximation $D_\text{in} / D_0 = \exp ( - \phi_\text{m} )$ valid for low $\epsilon_\text{mp}$ and $\phi_\text{m}$, and the scaling $D_\text{in} / D_0 \sim \phi_\text{m}^{-2/3}$ valid for high $\epsilon_\text{mp}$ and low $\phi_\text{m}$.
(f)~$\mathcal{P}(\phi_\text{m})/D_{0}$ at different $\epsilon_\text{mp}$.}\label{fig:KDP_FCC}
\end{figure*}

\textit{Result and discussion.}---
First we discuss the results from the SC lattice membranes.
\Figure~\ref{fig:KDP_SC}(a) shows the partitioning versus the LJ interaction energy, $\mathcal{K}(\epsilon_\text{mp})$, at various membrane packing $\phi_\text{m}$. The simulation results (symbols) reproduce very well the exact theoretical prediction (solid lines) in \Eq~\eqref{eq:K}. From  $\epsilon_\text{mp} = 0$ to non-zero interaction energies, $\mathcal{K}$ features a small jump to values smaller than unity because the excluded-volume of the membrane sites is switched on. For increasing $\epsilon_\text{mp}$, i.e., increasing attraction, $\mathcal{K}$ strongly rises exponentially as expected~\cite{SI}. The scaling with $\mathcal{K}\propto \exp(\beta \epsilon_{\rm mp})$ (dashed line), fits well the moderate packing fractions between 0.1 and 0.4 and is given as a guide. 

\Figure~\ref{fig:KDP_SC}(b) shows the scaled penetrant diffusivity $D_\text{in}(\epsilon_\text{mp})/D_0$ inside the membrane at different $\phi_\text{m}$. The diffusion is as expected always slower than bulk diffusion, $D_0$, due to crowding and diminishes monotonically with increasing $\epsilon_\text{mp}$. A Kramers' type scaling for activated diffusion~\cite{masaro1999physical}, $D_\text{in} \propto e^{-\beta \epsilon_\text{mp}}$,   fits the data for moderate packing and large attractions well.  
However, overlapping (many-body) potentials smoothen the energy landscape \cite{SI} and diffusion gets faster again~\cite{Putzel2014a, Ghosh2014a} for very dense SC membranes ($\phi_\text{m} \gtrsim 0.4$) but with a weaker scaling with $\epsilon_\text{mp}$. We computed the landscape roughness defined by the variance of partitioning $\sigma_\mu^2 = \overline{(e^{-\beta H_{\rm mp}(\mathbf r)} - \mathcal{K})^2}$, see \cite{SI}.

The permeability, the product of $\mathcal{K}$ and $D_{\rm in}$, now results from drastic cancellations in a non-trivial way.  $\mathcal{P}(\epsilon_\text{mp})/D_{0}$, shown in \Figure~\ref{fig:KDP_SC}(c), varies dramatically with $\phi_\text{m}$:  
For less crowded membranes ($\phi_\text{m} \lesssim 0.3$), decreasing diffusivity wins over increasing partitioning, thus permeability monotonically decreases. For intermediate membrane packing around $\phi_\text{m} = 0.3$, both diffusivity and partitioning exponentially grow or decay (see the dashed lines in panels (a) and (b)) and mostly cancel out, yielding~$\mathcal{P}$ around unity.
For highly crowded membranes ($\phi_\text{m} \gtrsim 0.3$), permeability is minimized first with respect to $\epsilon_\text{mp}$, and as $\epsilon_\text{mp}$ further increases, partitioning dominates over diffusivity, resulting in an exponential increase. 

\begin{figure*}[t]
\centering
\includegraphics[width = 1\textwidth]{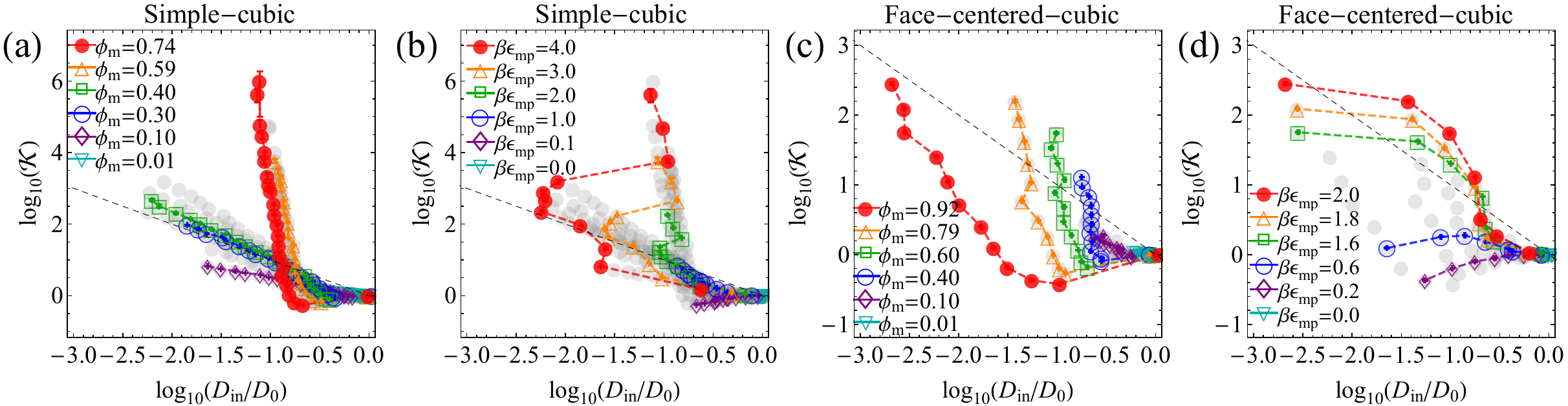}
\caption{Partitioning--diffusivity ($\mathcal{K}$--$D_\text{in}$) correlation diagram.
The gray symbols depict all the simulation results and colored symbols correspond to chosen parameters in the legends.
(a)~Simple-cubic (SC) membrane--penetrant systems at different membrane volume fractions $\phi_\text{m}$. 
(b)~SC membrane--penetrant systems at at different membrane--penetrant interactions $\epsilon_\text{mp}$.
(c)~Face-centered-cubic (FCC) membrane--penetrant systems at different $\phi_\text{m}$.
(d)~FCC membrane--penetrant systems at different $\epsilon_\text{mp}$.
The black dashed lines depict the equi-permeability line of $\mathcal{P}/D_{0}=1$.
}\label{fig:KD}
\end{figure*}

In \Fig~\ref{fig:KDP_SC}(d) partitioning versus packing fraction, $\mathcal{K}(\phi_\text{m})$, is shown (symbols), in excellent agreement with the relation in \Eq~\eqref{eq:K} (solid lines). As $\epsilon_\text{mp}$ varies from repulsive to attractive interactions, accordingly $\mathcal{K}(\phi_\text{m})$ changes from decreasing to increasing functions. The dashed line depicts a leading order approximation on a two-body level, $\mathcal{K}_\text{B2}(\phi_\text{m},\epsilon_\text{mp})=\exp\left[-2 c_\text{m} B_2^\text{mp}\right]$ for $\beta\epsilon_\text{mp}=1$, where $c_\text{m} \propto \phi_\text{m}$ is the membrane number density, and $B_2^\text{mp}$ is the second virial coefficient. The scaling agrees well with the simulation result for a wide range of densities~\cite{SI}. 

\Figure~\ref{fig:KDP_SC}(e) shows diffusivity versus packing fraction, $D_\text{in}(\phi_\text{m}) / D_0$. For large $\beta\epsilon_\text{mp} > 1.0$ and around $\phi_\text{m} = 0.4$, interestingly, $D_\text{in}$ is markedly minimized but we find also a local maximum around $\phi_\text{m} = 0.6$ exemplifying the competitive effects of smoothening the energy landscape (cf. Fig. S10 in~\cite{SI}) due to overlapping potentials and increasing steric constraints. 
The upper dashed line depicts the limiting law, $D_\text{in} / D_0 = \exp (- \phi_\text{m} )$,  based on the well known volume-exclusion ansatz~\cite{HAUS1987,masaro1999physical, Amsden1998, Ghosh2014a, lyd}, indeed found to be valid for low $\epsilon_\text{mp}$ (mostly repulsive interactions).  For high $\epsilon_\text{mp}$ and low $\phi_\text{m}$, diffusivity follows rather the power law $D_\text{in} / D_0 \sim \phi_\text{m}^{-2/3}$, limited by the Kramers' escape from a well and hopping to a neighboring well in the distance $l\sim\phi^{-1/3}$, and therefore $D_{\rm in} \sim l^2/\tau \sim \phi_c^{-2/3}$. The resulting permeability presented in \Fig~\ref{fig:KDP_SC}(f), exhibits again interesting features: It is minimized at $\phi_\text{m} \simeq 0.1$ for attractive membranes and then increases with packing (apart from the essentially repulsive case $\beta\epsilon_\text{mp}=0.1$). 
There is, on the other hand, indication of slight maximization and sharp decrease of $\mathcal{P}$ when approaching  $\phi_\text{m}=1$, due to vanishing partitioning in the impenetrable full packing limit.

The results change substantially when slightly varying the geometry of the membrane, thereby reshaping the underlying potential landscape roughness $\sigma_\mu$~\cite{SI}. The partitioning $\mathcal{K}(\epsilon_\text{mp})$ in the FCC lattice membrane at different $\phi_\text{m}$ are shown in \Fig~\ref{fig:KDP_FCC}(a). Partitioning again increases exponentially. The diffusivity $D_\text{in}(\epsilon_\text{mp})$ at different $\phi_\text{m}$ is shown in \Fig~\ref{fig:KDP_FCC}(b), where, unlike in the SC case, it decays more rapidly as $\phi_\text{m}$ increases, reflecting the strong effect of the membrane geometry.
As a striking consequence we find in \Fig~\ref{fig:KDP_FCC}(c) that permeability is markedly minimized. 

\Figure~\ref{fig:KDP_FCC}(d) shows $\mathcal{K}(\phi_\text{m})$ for various $\epsilon_\text{mp}$. For repulsive interactions, partitioning monotonically decreases as the sites pack more, driven by exclusion. For intermediate attractive interactions around $\beta\epsilon_\text{mp} = 0.6$, partitioning is maximized at an optimal packing around $\phi_\text{m} = 0.6$, resulting from a balance between the attraction and exclusion, as also found for model membranes of polymer networks~\cite{monchoPCCP2018}.  For highly attractive interactions, the maximum point of partitioning shifts towards the extreme overlapping regime $\phi_\text{m} \lesssim 1$. The diffusivity $D_\text{in}(\phi_\text{m}) / D_0$ is shown in \Fig~\ref{fig:KDP_FCC}(e), where the limiting laws (dashed lines) qualitatively embrace the simulation results. Finally, we show in \Fig~\ref{fig:KDP_FCC}(f) permeability $\mathcal{P}(\phi_\text{m})/D_{0}$ at different $\epsilon_\text{mp}$. When the system is highly attractive while densely and smoothly packed, the permeability is clearly maximized considerably before the packing reaches 100\%.

To better visualize the correlations and cancellations between partitioning and diffusivity, we plot $\mathcal{K}$ versus $D_\text{in}$ diagrams in \Fig~\ref{fig:KD}, where the gray symbols depict all the simulation data, and the black dashed lines depict the equi-permeability line of $\mathcal{P}/D_{0}=1$, where the contributions of $\mathcal{K}$ and $D_\text{in}$ exactly cancel. We observe clear anti-correlations along the equi-permeability line, that is, in general partitioning and diffusion like to cancel out. In other words, increasing attraction slows down mobility in a similar, exponential fashion. However, depending on the potential details, in some cases the diagram shows more complex pathways (dashed lines between the symbols) in the $\mathcal{K}$--$D_\text{in}$ phase space.  For instance, for the SC sites $\mathcal{K}$ becomes less sensitive on $D_\text{in}$ but the magnitude changes over 6 decades when they are highly dense (panel (a)), and $D_\text{in}$ is significantly minimized when they are highly attractive (panel (b)), pointing to very smooth potential landscapes.  For FCC sites, the diagram clearly shows global minimization of $\mathcal{K}$ over 3 decades of $D_\text{in}$ (panel (c)) and permeability maximization (panel (d)).

The theoretical description for diffusivity in our work is limited to scaling theories.  The description of diffusion in multi-dimensional energy landscape, even for non-interacting penetrants, is very complex, see, e.g., Refs.~\cite{masaro1999physical, Ben2}, and no explicit or unified analytic framework is available.  For dense membranes, however, we note that we attempted to adopt the excess entropy scaling approach, $D_\text{in} \sim D_{0} e^{b\Delta S/ \kB}$~\cite{Rosenfeld1977,Dzugutov1996,seki2015relationship}. We find qualitative agreement of the theory with the simulation results~\cite{SI}. Also, we note that we tested non-ideal penetrants with non-vanishing excluded volume in the SC membrane system and we found the  same qualitative features as for the ideal penetrants~\cite{SI}. 

Within the solution-diffusion model the final permeability can be conveniently interpreted by the individual or combined action of two intuitive processes, the partitioning and the mobility of the solutes. The maximum in permeability for example can then be traced back to microscopic phenomena, such as excluded volume or smoothened energy landscapes. This in-depth interpretation may lead to improved design rules for membrane manufacture~\cite{Baker,chauhan,Thomas2001transport,Ulbricht2006,Baker2014}.
Interestingly, our apparently simple, very ordered systems behave very complex (SC versus FCC), much owed to the periodicity of the potential energy landscapes.
\textcolor{black}{In reality, membranes will have some amount of disorder that may smear out some effects; however, we do not observe less complex behavior in a more disordered array of dense attractive sites~\cite{SI}.}
A recent paper, however, demonstrated that the permeability of a polymer membrane increased by orders of magnitude when the polymer crystallizes and is more ordered~\cite{van2018increasing}. The amount of order therefore may be in principle an important tuning parameter.

In summary, we demonstrated how to tune the permeability \textcolor{black}{(in the overdamped regime)} of dense membranes over orders of magnitude by shaping nanoscale potentials.
The complex behavior of the permeability results from a strong anti-correlation and partial cancellation between penetrant partitioning and diffusivity, particularly in highly attractive membranes and fine-tuned by details of the potential landscape. This interaction-specific control of membrane permeation bears possible rational design applications in material science and nanofluidics to selectively transport solvents and solutes for the desired material function.
High resolution 3D laser micro- and nanoprinting with a variety of materials has become possible~\cite{barner20173d} so that our results shall be useful for membrane design with sub-micron internal structure, to control the architecture, pore shape, porosity, or interconnectivity of the scaffold, enhancing the membrane design~\cite{lee2016potential} or tissue engineering~\cite{lee20153d} with 3D printing technology.

\begin{acknowledgments}
The authors thank Matthias Ballauff and Benjamin Rotenberg for fruitful discussions. This project has received funding from the European Research Council (ERC) under the European Union's Horizon 2020 research and innovation programme (grant agreement Nr.~646659).
MK acknowledges the financial support from the Slovenian Research Agency (research core funding no.~P1-0055).
The simulations were performed with resources provided by the North-German Supercomputing Alliance (HLRN).
\end{acknowledgments}


\end{document}